\documentclass[preprint,tightenlines,aps]{revtex4}
\begin{document}
\title{A multiquark description of the $D_{sJ}(2860)$ and $D_{sJ}(2700)$}
\author{J. Vijande$^{1}$, A. Valcarce$^{2}$, F. Fern\'{a}ndez$^{3}$.}
\affiliation{$^{1}$Dpto. de F\' \i sica At\'omica, Molecular y Nuclear,
Universidad de Valencia - CSIC, E-46100 Burjassot, Valencia, Spain.\\
$^{2}$Departamento de F\'\i sica Fundamental,
Universidad de Salamanca, E-37008 Salamanca,
Spain.\\
$^{3}$Grupo de F\'{\i}sica Nuclear and IUFFyM,
Universidad de Salamanca, E-37008 Salamanca,
Spain}
\vspace*{1cm} 
\begin{abstract}
Within a theoretical framework that accounts for all open-charm mesons, 
including the $D_0^*(2308)$, the $D_{sJ}^*(2317)$ and the $D_{sJ}(2460)$, 
we analyze the structure and explore possible quantum number assignments 
for the $D_{sJ}(2860)$ and the $D_{sJ}(2700)$ mesons reported 
by BABAR and Belle Collaborations. 
The open-charm sector is properly described if considered as a combination 
of conventional quark-antiquark states and four--quark components. All negative 
parity and $2^+$ states can be understood in terms only of $q\bar q$ components, 
however the description of the $0^+$ and $1^+$ mesons is improved whenever the 
mixing between two-- and four--quarks configurations is included. We analyze all 
possible quantum number assignments for the $D_{sJ}(2860)$ in terms of 
both $c\bar s$ and $cn\bar s\bar n$ configurations. 
We discuss the role played by the electromagnetic and strong decay widths as 
basic tools to distinguish among them. The broad structure reported by BABAR 
near 2.7 GeV is also analyzed.

\vspace*{2cm} 
\newline
\noindent Pacs: 14.40.Lb, 14.40.Ev, 12.39.Pn. 
\end{abstract}
\maketitle
\newpage

The open-charm sector does not cease to amaze both theorists and experimentalists
with new states that defy our understanding of the heavy-meson spectra.
Two new mesons have recently joined the open-charm zoo.
BABAR Collaboration reported the observation of a new $D_s$ state,
the $D_{sJ}(2860)$, with a mass of 2856.6$\pm1.5\pm5.0$ MeV and a width of 
48$\pm7\pm10$ MeV in the analysis of the $DK$ spectra. No structures seem to appear
in the $D^*K$ invariant mass distribution in the same range of masses. This state was reported 
together with a broad enhancement near 2.7 GeV with a tentative mass of 
2688$\pm4\pm2$ MeV and a width $112\pm7\pm36$ MeV \cite{Bab06}. Since the only 
reported decay channels observed by BABAR correspond to two pseudoscalar mesons, 
the assignment of natural parity quantum numbers, $J^P=0^+,1^-,2^+,3^-....$, 
is strongly favored. The second state, the $D_{sJ}(2700)$, was observed by Belle 
Collaboration in the decay $B^+\to\overline{D^0}D^0K^+$ with a mass of 2708$\pm$9$^{+11}_{-10}$
MeV, a width of 108$\pm$23$^{+30}_{-31}$ MeV, and quantum numbers $J^{P}=1^-$~\cite{Bro07}.
The $D_{sJ}(2860)$ is not observed in the Belle data.

The $D_{sJ}(2860)$ and $D_{sJ}(2700)$ mesons are new members of a long list of charm 
resonances reported during the last few years. In 2003 BABAR reported the observation 
of a charm-strange state, the $D_{sJ}^*(2317)$, with a mass of 2316.8$\pm$0.4$\pm$3 MeV 
and a width of less than $4.6$ MeV \cite{Bab03}. This state was soon after confirmed by 
CLEO \cite{Cle03} and Belle Collaborations \cite{Bel04}. Besides, BABAR also pointed out 
the existence of another charm-strange meson, the $D_{sJ}(2460)$ \cite{Bab03}. This 
resonance was measured by CLEO \cite{Cle03} and confirmed by Belle \cite{Bel04} with a 
mass of 2457.2$\pm$1.6$\pm$1.3 MeV and a width less than $5.5$ MeV. Belle results are 
consistent with the spin-parity assignments of $J^P=0^+$ for the $D^*_{sJ}(2317)$ and 
$J^P=1^+$ for the $D_{sJ}(2460)$. Thus, these two states are definitively well established 
and confirmed independently by different experiments. In the nonstrange sector Belle 
reported the observation of a nonstrange broad scalar resonance named $D^*_0$ with a mass 
of $2308\pm 17\pm 15\pm 28$ MeV and a width $276 \pm 21 \pm 18 \pm 60$ MeV \cite{Belb4}. 
A state with similar properties has been suggested by FOCUS Collaboration \cite{Foc04} 
during the measurement of excited charm mesons $D^*_2$. SELEX Collaboration at Fermilab 
\cite{Sel04} has reported an state with a mass of $2632.5\pm 1.7$ MeV and a width 
smaller than 17 MeV. However, up to now no other experiment has been able to confirm
the existence of this resonance \cite{Bab04,Bab06}.

The positive parity open-charm mesons present unexpected properties quite different 
from those predicted by quark potential models if a pure $c\bar q$ configuration 
is considered. If they would correspond to standard $P-$wave mesons made of a 
charm quark and a light antiquark their masses would be larger, around 2.48 GeV for 
the $D_{sJ}^*(2317)$, 2.55 GeV for the $D_{sJ}(2460)$, and 2.46 GeV for the $D^*_0(2308)$. 
They would be therefore above the $DK$, $D^*K$, and $D\pi$ thresholds respectively, 
being broad resonances. However, as stated above, the $D_{sJ}^*(2317)$ and 
$D_{sJ}(2460)$ are very narrow. In the case of the $D^*_0(2308)$ the large width 
observed would be expected but not its low mass.
Although there are several theoretical interpretations for the masses and 
widths of some of the positive parity states  $D^*_0(2308)$, $D_{sJ}^*(2317)$, 
and $D_{sJ}(2460)$ \cite{Swa06}, in Ref. \cite{Vij06} it was shown that the 
difficulties to identify the three of them with conventional $c\overline{q}$ 
mesons are rather similar to those appearing in the light-scalar meson sector and 
may be indicating that other configurations, as for example four--quark 
components, may be playing a role. $q\overline{q}$ states are more easily identified 
with physical hadrons when virtual quark loops are not important. This is the case 
of the pseudoscalar and vector mesons, mainly due to the $P-$wave nature of this 
hadronic dressing. On the contrary, in the scalar sector it is the $q\overline{q}$ pair 
the one in a $P-$wave state, whereas quark loops may be in a $S-$wave. In this case 
the intermediate hadronic states that are created may play a crucial role in the 
composition of the resonance, in other words unquenching may be important. The 
vicinity of these components to the lightest $q\bar q$ state implies that they 
have to be considered. This has been shown as a possible interpretation of the 
low-lying light-scalar mesons, where the coupling of the scalar $q\overline{q}$ 
nonet to the lightest $qq\bar{q} \bar{q}$ configurations allows for an almost 
one-to-one correspondence between theoretical states and experiment \cite{Vij05}. 
The possible role played by non$-q\bar q$ components in the description of the
$D_{sJ}(2860)$ was illustrated in Ref. \cite{Bev06}. In this work it was proposed
that this state could be understood within a unitarized meson model as a quasi-bound $c\overline{s}$ 
state coupled with the
nearby $S-$wave $DK$ threshold, therefore 
being the first radial excitation of the $D_{sJ}^*(2317)$.

In non-relativistic quark models gluon degrees 
of freedom are frozen and therefore the wave 
function of a zero baryon number (B=0) hadron may be written as
\begin{equation}
\label{mes-w}
\left|\rm{B=0}\right>=\Omega_1\left|q\bar q\right>+\Omega_2\left|qq\bar q \bar q\right>+....
\end{equation}
where $q$ stands for quark degrees of 
freedom and the coefficients $\Omega_i$ take into 
account the mixing of two-- and four--quark states.
$\left|\rm{B=0}\right>$ systems could then be 
described in terms of a hamiltonian 
\begin{eqnarray}
&&H=H_0 + H_1 \,\,\,\,\, {\rm being} \,\,\,\,\, 
H_0 = \left( \matrix{H_{q\bar q} & 0 \cr 0 & H_{qq\bar q\bar q} \cr } \right) \nonumber\\
&&{\rm and}\,\,\,\,H_1 = \left( \matrix{0 & V_{q\bar q \leftrightarrow qq\bar q\bar q} \cr 
V_{q\bar q \leftrightarrow qq\bar q\bar q} & 0 \cr } \right)\,,
\label{eq1}
\end{eqnarray}
where $H_0$ is a constituent quark model hamiltonian described below and
$H_{1}$, that takes into account the mixing between
$q\overline{q}$ and $qq\bar{q}\bar{q}$ configurations, includes the
annihilation operator of a quark-antiquark pair into the vacuum. This
operator could be described using the $^{3}P_{0}$ model, however, since this 
model depends on the vertex parameter, we prefer in a
first approximation to parametrize this coefficient 
by looking to the quark pair that is annihilated and 
not to the spectator quarks that will form the final 
$q\overline{q}$ state. Therefore we have taken 
$V_{q\overline{q} \leftrightarrow qq\bar{q}\bar{q}}=\gamma $.
If this coupling is weak enough one can solve independently 
the eigenproblem for the hamiltonians $H_{q\overline{q}}$ 
and $H_{qq\bar{q}\bar{q}}$, treating $H_{1}$ perturbatively. 
The two--body problem has been solved exactly by means of the Numerov
algorithm \cite{Vijb5}. The four--body problem has been solved
by means of a variational method using the most general 
combination of gaussians as trial
wave functions \cite{Suz98}. 
In particular, the so-called {\it mixed terms} (mixing the various
Jacobi coordinates) that are known to have a great influence in
the light quark case have been considered. 

It is our purpose in this work to use a standard constituent quark model that provides
with a good description of the meson and baryon spectra
and also the baryon-baryon phenomenology for the description
of the open-charm mesons.
For this purpose, we will address the study of hadrons with
zero baryon number described as clusters of quarks confined
by a realistic interaction.
The model is based on the assumption
that the constituent quark mass appears because of the spontaneous
breaking of the original $SU(3)_{L}\otimes SU(3)_{R}$ chiral symmetry at
some momentum scale. As a consequence of such a symmetry breaking,
quarks acquire a constituent mass and Goldstone bosons
are exchanged between the quarks.
Beyond the chiral symmetry breaking scale, one expects the dynamics to
be governed by QCD perturbative effects, that are taken into account
through the one-gluon-exchange potential.
Finally, any model imitating QCD should incorporate another
nonperturbative effect, confinement. It remains an unsolved problem
to derive confinement from QCD in an analytic manner. The only indication
we have on the nature of confinement is through lattice studies,
showing that $q\bar q$ systems are well reproduced at short
distances by a linear potential. Such potential can be physically
interpreted in a picture in which the quark and the antiquark are linked
with a one-dimensional color flux tube. The spontaneous creation of
light-quark pairs may give rise to a breakup of the color flux tube,
what has been proposed that translates into a screened
potential \cite{Bal01}, in such a way that the potential
saturates at some interquark distance.
Explicit expressions of the interacting $qq$ and $q \bar q$ potentials and a
more detailed description of the model can be found in Refs.~\cite{Vijb5,Vij05}
where the various parameters are given. 

\begin{table}
\caption{$c\overline s$ masses (QM), in MeV, below 3 GeV.
Experimental data (Exp.) are taken from
Ref. \protect\cite{Eid04}.}
\label{t1}
\begin{center}
\begin{tabular}{|cc|ccc|}
\hline
&$nL$ $J^P$&State		&QM 	& Exp.		\\
\hline
&$1S$ $0^-$&$D_{s}$ 	  	& 1981  &1968.49$\pm$0.34	\\
&$2S$ $0^-$&$-$ 	  	& 2699  &$-$	\\
&$1S$ $1^-$&$D^*_{s}$	  	& 2112  &2112.3$\pm$0.5	\\
&$2S$ $1^-$&$-$		  	& 2764  &$-$	\\
&$1P$ $0^+$&$D^*_{sJ}(2317)$	& 2489  &2317.8$\pm$0.6	\\
&$2P$ $0^+$&$-$			& 2966  &$-$	\\
&$1P$ $1^+$&$D_{sJ}(2460)$	& 2578  &2459.6$\pm$0.6	\\
&$1P$ $1^+$&$D_{s1}(2536)$	& 2543  &$2535.35 \pm 0.34 \pm 0.5$	\\
&$1P$ $2^+$&$D_{s2}(2573)$	& 2582  &2572.6$\pm$0.9	\\
&$1D$ $1^-$&$-$			& 2873  &$-$	\\
&$1D$ $2^-$&$-$			& 2883  &$-$	\\
&$1D$ $3^-$&$-$			& 2882  &$-$	\\
\hline
\end{tabular}
\end{center}
\end{table}

A thoroughly study of the full meson spectra has been presented in Ref. \cite{Vijb5}, 
with special attention in Ref. \cite{Vij06} to the open-charm sector. Using this model we have calculated
the $c\bar s$ masses up to 3 GeV listed in Table \ref{t1}. It can be seen how the open-charm states are easily identified with
standard $c \overline{q}$ mesons except for the cases of the $D_{sJ}^*(2317)$ and the $D_{sJ}(2460)$.
This behavior is shared by almost all quark potential model calculations \cite{Swa06}. 
Although the situation from lattice QCD is far from being definitively established,
similar difficulties have been observed both in quenched and unquenched approaches \cite{Lat06}.
The same conclusion may also be drawn from heavy quark symmetry arguments. Within this approach 
the scalar $c\overline s$ state belongs to the $j=1/2$ doublet, 
but since the $j=3/2$ doublet is identified with the
narrow $D_{s2}(2573)$ and $D_{s1}(2536)$ (with total widths 
of 15$^{+5}_{-4}$ MeV and $<2.3$ MeV, respectively) the scalar state is
expected to have a much larger width than the one measured for the
$D_{sJ}^*(2317)$ \cite{WI90}. Thus, one possibility for these states beyond the 
naive $q\bar q$ assignment is to interpret them as four--quark resonances within 
the quark model. The results obtained for the $cn\bar s\bar n$ 
and $cn\bar n\bar n$ configurations in Ref. \cite{Vij06} using the constituent 
quark model outlined above are shown in Table \ref{t2}. It can be seen that the 
$I=1$ and $I=0$ states obtained are far above the corresponding 
strong decay threshold and therefore should be broad, what rules out a pure four--quark 
interpretation of the positive-parity open-charm mesons. 

\begin{table}
\caption{$cn\bar s\bar n$ and $cn\bar n\bar n$ masses, in MeV.}
\label{t2}
\begin{center}
\begin{tabular}{|cc|cc|c|}
\hline
\multicolumn{4}{|c|} {$cn\bar s\bar n$} & $cn\bar n\bar n$ \\
\hline
\multicolumn{2}{|c|} {$J^{P}=0^{+}$} & \multicolumn{2}{|c|}{$J^{P}=1^{+}$}&$J^{P}=0^{+}$ \\
$I=0$& $I=1$ &  $ I=0$& $I=1$  & $I=1/2$\\
\hline
2731&  2699& 2841&2793 &2505\\
\hline
\end{tabular}
\end{center}
\end{table}

As discussed above, for $P-$wave mesons the hadronic dressing is in a $S-$wave,
thus physical states may correspond to a mixing of 
two-- and four--body configurations, Eq. (\ref{mes-w}). 
In the isoscalar sector, the 
$cn\bar s \bar n$ and $c\bar s$ states get mixed, as it happens
with $cn \bar n\bar n$ and $c\bar n$ for the $I=1/2$ case. 
The parameter $\gamma$ was fixed in Ref. \cite{Vij06} to reproduce the 
mass of the $D_{sJ}^*(2317)$ meson, being $\gamma=240$ MeV. 
The results obtained are shown in Table \ref{t3}. From these 
results one can appreciate that the description of the positive parity open-charm 
mesons improves when four--quark components are considered.

With respect to the new resonance reported by BABAR, it can be see from Tables \ref{t1} 
and \ref{t3} that among all possibilities only three states are close to its experimental 
mass, 2856.6$\pm1.5\pm5.0$ MeV. They correspond to the $0^+$ $c\bar s+cn\bar s\bar n$ 
(45\% and 55\% probability respectively) excitation and the $1^-$ and $3^-$ $c\bar s$ $D-$waves, 
being their energies 2847, 2873, and 2882 MeV. All other possibilities, 
like for instance the $2S$ $1^-$ or the $2P$ $2^+$, are more than 100 MeV above or 
below the experimental energy. The $2S$ $1^-$ excitation obtained within our model
with an energy of 2764 MeV is a good candidate to be identified with the $D_{sJ}(2700)$
reported by Belle. Concerning the broad bump reported by BABAR at 2.7 GeV, 
if different from the $D_{sJ}(2700)$, two states appear 
as possible candidates, the $2S$ $0^-$ radial $c\bar s$ excitations and the 
isovector $0^+$ $cn\bar s\bar n$ ground state, both of them with a mass of 2699 MeV.

\begin{table}
\caption{Probabilities (P), in \%, of the wave function components 
and masses (QM), in MeV, of the open-charm mesons with $I=0$ (left) and
$I=1/2$ (right) once the mixing between $q\bar q$ and $qq\bar q\bar q$ configurations 
is considered. Experimental data (Exp.) are taken from Ref. \cite{Eid04}
for $I=0$ and from Ref. \cite{Belb4} for $I=1/2$.} 
\label{t3}
\begin{center}
\begin{tabular}{|c|cc||c|cc||c|cc|}
\hline
\multicolumn{6}{|c||}{$I=0$}    & \multicolumn{3}{|c|}{$I=1/2$} \\
\hline
\multicolumn{3}{|c||}{$J^P=0^+$}    & \multicolumn{3}{|c||}{$J^P=1^+$} &
\multicolumn{3}{|c|}{$J^P=0^+$} \\
\hline
QM                  	&2339   	&2847	&QM			&2421  		&2555 &
QM                   	&2241 				&2713    \\
Exp.                	&2317.8$\pm$0.6	&$-$	&Exp.			&2459.6$\pm$0.6	&$2535.35 \pm 0.34 \pm 0.5$ &
Exp.			&2308$\pm$17$\pm$32	&$-$\\
\hline
P($cn\bar s\bar n$) 	&28   		&55	&P($cn\bar s\bar n$)	&25  		&$\sim 1$ &
P($cn\bar n\bar n$)  	&46        			&49  \\
P($c\bar s_{1^3P}$) &71   &25  &P($c\bar s_{1^1P}$)	&74  &$\sim 1$ 	&
P($c\bar n_{1P}$)    	&53        			&46 \\
P($c\bar s_{2^3P}$) &$\sim 1$  &20  &P($c\bar s_{1^3P}$)&$\sim 1$ &98	&
P($c\bar n_{2P}$)    	&$\sim 1$  			&5 \\
\hline
\end{tabular}
\end{center}
\end{table}

From the analysis of the masses alone it is not possible to distinguish among all candidates 
for the new resonances. However, the structure of the $D^*_{sJ}$ mesons could be scrutinizedy 
apart from their masses, also through the study of their decay widths. The strong decay width
of a hypothetical $J^P=0^+$ $D_{sJ}(2860)$, either $c\bar s$ or $cn\bar s\bar n+c\bar s$,
 into $D^*\,K$ or $D\,K^*$ is forbidden due to quantum number conservation. This is consistent
with the absence of $D\,K^*$ or $D^*\,K$ signals in the experimental data~\cite{Bab06}. The
ratio $\Gamma[D_{sJ}(2860)\to D^*K]/\Gamma[D_{sJ}(2860)\to DK]$ has been studied for several
different quantum numbers using the $^3P_0$ model \cite{Zha07} and arguments based on heavy
quark expansions \cite{Col07}. Some of their results are quoted in Table~\ref{tabstrong}. A
possible assignment $L=2$ $J^P=1^-$ would result in a too large $\Gamma[D_{sJ}(2860)\to DK]$
decay width, although very different values are obtained depending on the approach considered, 
while the $L=2$ $J^P=3^-$ $\Gamma[D_{sJ}(2860)\to D^*K]/\Gamma[D_{sJ}(2860)\to DK]$ ratio
seems to indicate a sizable $\Gamma[D_{sJ}(2860)\to D^*K]$ decay width that has not yet been
observed. Concerning the electromagnetic decay widths of the $J^P=0^+$ candidate,
the formalism necessary to study the $\Gamma[c\bar s\to c \bar s+\gamma]$, 
$\Gamma[cn\bar s\bar n\to cn \bar s\bar n+\gamma$], and $\Gamma[cn\bar s\bar n\to c \bar s+\gamma]$ 
processes has been described in detail in Ref. \cite{Vij06}, where it was proposed that the 
ratio $R={D_{sJ}(2460)\to D^{+}_s\gamma}/{D_{sJ}(2460)\to D^{*+}_s\gamma}$ could be an important 
tool to distinguish between a $q\bar q$ structure ($R\approx 1$) or a $qq\bar q\bar q+q\bar q$ one 
($R\approx 100$) for the open-charm mesons. Following this formalism one obtains for the electromagnetic decay 
$\Gamma[D_{sJ}(2860)\,[0^+] \to D_s^*\,\gamma]$=13.67 keV. At the same time, if a pure $1P$ $c\bar s$ 
structure is assumed for the $D_{sJ}(2317)$ then the $0^+$ $D_{sJ}(2860)$ should correspond to a $2P$ 
excitation. One can also evaluate this decay obtaining a much smaller value, 
$\Gamma[D_{sJ}(2860)\,0^+ \to D_s^*\,\gamma]$=1.8 eV, due to the presence of a node in the $2P$ 
wave function. Therefore, the mixed scenario would produce a sizable value for the electromagnetic 
decay width $\Gamma[D_{sJ}(2860)\to D_s^*\,\gamma]$ only if an scalar state with a dominant 
four--quark component is present.

\begin{table}
\caption{$\Gamma[D_{sJ}(2860)\to D^*K]/\Gamma[D_{sJ}(2860)\to DK]$ ratio and
$\Gamma[D_{sJ}(2860)\to DK]$ strong decay width in MeV for different approaches.}
\label{tabstrong}
\begin{center}
\begin{tabular}{|c|cc|cc|}
\hline
$J^P$	& \multicolumn{2}{|c|}{${{\Gamma[D_{sJ}(2860)\to D^*K]}\over{\Gamma[D_{sJ}(2860)\to DK]}}$}&
\multicolumn{2}{|c|}{$\Gamma[D_{sJ}(2860)\to DK]$}\\
     	& \multicolumn{2}{|c|}{Exp: Not observed}	& \multicolumn{2}{|c|}{Exp: 48$\pm7\pm10$ MeV}\\
\cline{2-5}
	& \cite{Zha07}	& \cite{Col07}	& \cite{Zha07}	& \cite{Col07} \\
\hline
$1^-$	& 0.17		& 0.06		& 84		& $>$1000	\\
$3^-$	& 0.59		& 0.37		& 22		& Narrow \\
\hline
\end{tabular}
\end{center}
\end{table}

If the $D_{sJ}(2860)$ is definitively confirmed as a scalar meson, this will point to the
existence of a non-strange partner with an energy of 2713 MeV and an important four--quark 
component (49\%), being this result in the same line as the one reported in Ref.\cite{Bev06}. 
Furthemore, an isovector $1^+$ $cn\bar s\bar n$ state with a mass of 2793 Mev is also predicted.

The interpretation we have just presented for the new resonances measured by BABAR and Belle
within a formalism that includes the mixing of two-- and four--quark states has also been used to account for the 
other experimentally observed open-charmed states and also for the light-scalar mesons within 
the same constituent quark model \cite{Vij05,Vij06}. It is therefore the first time that a 
coherent analysis of all known states within the meson spectra in terms of two-- and four--quark 
states is performed, what gives us confidence on the mechanism proposed. Nonetheless, one should 
not forget that in the literature there is a wide variety of interpretations for the open-charm mesons. 
Therefore, the final answer could only be obtained from precise experimental data that would allow 
to discriminate between the predictions of different theoretical models \cite{Ams04}.

As a summary, we have obtained a rather satisfactory description of 
the open-charm mesons in terms of two-- and four--quark configurations, including the new states 
recently reported by BABAR and Belle. The mixing between these two components is responsible for the 
unexpected low mass and widths of the
$D_{sJ}^*(2317)$, $D_{sJ}(2460)$, and $D_0^*(2308)$ and also offers a possible interpretation for 
the $D_{sJ}(2860)$ as a scalar meson. The electromagnetic and strong decay widths give hints that would 
help in distinguishing the nature of these states. In particular, the study of the decays
$\Gamma[D_{sJ}(2860)\to D_s^*\,\gamma]$ and $\Gamma[D_{sJ}(2860)\to D^*\,K]$ are
ideally suited for this task. A clear signal for this electromagnetic decay mode together with 
the absence of the strong one would point to an scalar state with an involved structure in terms of two-- and 
four--quark components. 
The $1^-$ state at 2708 MeV observed by Belle can be interpreted 
as a $c\bar s$ 2S excitation whereas for the broad bump around 2.7 GeV reported by BABAR two candidates can be found, although 
more experimental data is needed before drawing any conclusion. We encourage experimentalists on 
the confirmation of the results reported by BABAR and Belle and on the measurement of the electromagnetic 
and strong decay widths of the open-charm positive parity states. Such a study would help to distinguish not 
only among the possible quantum numbers allowed for the new BABAR resonance, but also to clarify 
the exciting situation of the open-charm mesons and the role played by multiquark configurations 
in the meson spectra.

This work has been partially funded by Ministerio de Ciencia y Tecnolog\'{\i}a
under Contract No. FPA2007-65748, and by Junta de Castilla y Le\'{o}n
under Contract No. SA016A17.


\begin{thebibliography}{}
\bibitem{Bab06} BABAR Collaboration, B. Aubert {\it et al},
			Phys. Rev. Lett. {\bf 97}, 222001 (2006).

\bibitem{Bro07} Belle Collaboration, J. Brodzicka {\it et al.},
			Phys. Rev. Lett. {\bf 100}, 092001 (2008).

\bibitem{Bab03} BABAR Collaboration, B. Aubert {\it et al.}, 
			Phys. Rev. Lett. {\bf 90}, 242001 (2003).

\bibitem{Cle03} CLEO Collaboration, D. Besson {\it et al.}, 
			Phys. Rev. D {\bf 68}, 032002 (2003).

\bibitem{Bel04} Belle Collaboration, Y. Mikani {\it et al.}, 
			Phys. Rev. Lett. {\bf 92}, 012002 (2004).

\bibitem{Belb4} Belle Collaboration, K. Abe {\it et al.}, 
			Phys. Rev. D {\bf 69}, 112002 (2004).

\bibitem{Foc04} FOCUS Collaboration, J.M. Link {\it et al.}, 
			Phys. Lett. B {\bf 586}, 11 (2004).

\bibitem{Sel04} SELEX Collaboration, A.V. Evdokimov {\it et al.}, 
			Phys. Rev. Lett. {\bf 93}, 242001 (2004).

\bibitem{Bab04} BABAR Collaboration, B. Aubert {\it et al.}, 
			hep-ex/0408087. 
                Belle Collaboration, B. Yabsley,
			AIP Conf. Proc. {\bf 792}, 875 (2005).
                FOCUS Collaboration, R. Kutschke,
			E831-doc-701-v2.

\bibitem{Swa06} E.S. Swanson,
			Phys. Rep. {\bf 429}, 243 (2006) and references therein.

\bibitem{Vij06} J. Vijande, F. Fern\'andez, and A. Valcarce,
                Phys. Rev. D {\bf 73}, 034002 (2006).

\bibitem{Vij05} J. Vijande, A. Valcarce, F. Fern\'andez, and B. Silvestre-Brac,
                        Phys. Rev. D {\bf 72}, 034025 (2005).

\bibitem{Bev06} E. van Beveren and G. Rupp, 
		Phys. Rev. Lett. {\bf 97}, 202001 (2006).

\bibitem{Vijb5} J. Vijande, F. Fern\'andez, and A. Valcarce, 
			J. Phys. G {\bf 31}, 481 (2005).

\bibitem{Suz98} Y. Suzuki and K. Varga,
			Lecture Notes in Physics M {\bf 54}, 1 (1998);
		J. Vijande, F. Fern\'andez, A. Valcarce, and B. Silvestre-Brac,
			Eur. Phys. J. A {\bf 19} 383 (2004). 

\bibitem{Bal01} G.S. Bali, 
			Phys. Rep. {\bf 343}, 1 (2001).

\bibitem{Eid04} C. Amsler {\it et al.}, 
		Phys. Lett. B{\bf 667}, 1 (2008). 

\bibitem{Lat06} J. Hein {\it et al.}, 
			Phys. Rev. D {\bf 62}, 074503 (2000);
		G.S. Bali, 
			Phys. Rev. D {\bf 68}, 071501(R) (2003);
		UKQCD Collaboration, P. Boyle,
			Nucl. Phys. B (Proc. Supp.) {\bf 63}, 314 (1998);
				{\it et al.},
			Nucl. Phys. B (Proc. Supp.) {\bf 53}, 398 (1997);
		A. Dougall, {\it et al.}
			Phys. Lett. B {\bf 569}, 41 (2003).

\bibitem{WI90} N. Isgur and M. B. Wise,
		Phys. Lett. B {\bf 237}, 527 (1990); {\it ibid}
		Phys. Lett. B {\bf 232}, 113 (1989).

\bibitem{Zha07} B. Zhang, X. Liu, W.-Z. Deng, and S.-L. Zhu,
		Eur. Phys. J. C {\bf 50} 617 (2007). 

\bibitem{Col07} P. Colangelo, F. De Facio, R. Ferrandes, and S. Nicotri,
		Prog. Theor. Phys. Suppl. {\bf 168} 202 (2007);
 		P. Colangelo, F. De Fazio, and S. Nicotri, 
		Phys. Lett. B {\bf 642} 48 (2006). 

\bibitem{Ams04} C. Amsler and N.A. Tornqvist,
        Phys. Rep. {\bf 389}, 61 (2004) and references therein.

\end{thebibliography}
\end{document}